\documentclass[twoside,twocolumn,9pt]{article}
\usepackage{ caption, subcaption}
\usepackage{extsizes}
\usepackage[super,sort&compress,comma]{natbib}
\usepackage{bibentry} 
\usepackage[version=3]{mhchem}
\usepackage[left=1.5cm, right=1.5cm, top=1.785cm, bottom=2.0cm]{geometry}
\usepackage{balance}
\usepackage{times,mathptmx}
\usepackage{sectsty}
\usepackage{graphicx} 
\usepackage{lastpage}
\usepackage[format=plain,justification=justified,singlelinecheck=false,font={stretch=1.125,small,sf},labelfont=bf,labelsep=space]{caption}
\usepackage{float}
\usepackage{fancyhdr}
\usepackage{fnpos}
\usepackage[english]{babel}
\addto{\captionsenglish}{%
  
}
\usepackage{array}
\usepackage{droidsans}
\usepackage{charter}
\usepackage[T1]{fontenc}
\usepackage[usenames,dvipsnames]{xcolor}
\usepackage{setspace}
\usepackage[compact]{titlesec}
\usepackage{hyperref}

\usepackage{epstopdf}

\definecolor{cream}{RGB}{222,217,201}

\begin{document}
\pagestyle{plain}



\makeFNbottom
\makeatletter
\renewcommand\LARGE{\@setfontsize\LARGE{15pt}{17}}
\renewcommand\Large{\@setfontsize\Large{12pt}{14}}
\renewcommand\large{\@setfontsize\large{10pt}{12}}
\renewcommand\footnotesize{\@setfontsize\footnotesize{7pt}{10}}
\makeatother

\renewcommand{\thefootnote}{\fnsymbol{footnote}}
\renewcommand\footnoterule{\vspace*{1pt}%
\color{cream}\hrule width 3.5in height 0.4pt \color{black}\vspace*{5pt}} 
\setcounter{secnumdepth}{5}

\makeatletter 
\renewcommand\@biblabel[1]{#1}            
\renewcommand\@makefntext[1]%
{\noindent\makebox[0pt][r]{\@thefnmark\,}#1}
\makeatother 
\renewcommand{\figurename}{\small{Fig.}~}
\sectionfont{\sffamily\Large}
\subsectionfont{\normalsize}
\subsubsectionfont{\bf}
\setstretch{1.125} 
\setlength{\skip\footins}{0.8cm}
\setlength{\footnotesep}{0.25cm}
\setlength{\jot}{10pt}
\titlespacing*{\section}{0pt}{4pt}{4pt}
\titlespacing*{\subsection}{0pt}{15pt}{1pt}

\renewcommand{\headrulewidth}{0pt} 
\renewcommand{\footrulewidth}{0pt}
\setlength{\arrayrulewidth}{1pt}
\setlength{\columnsep}{6.5mm}
\setlength\bibsep{1pt}

\makeatletter 
\newlength{\figrulesep} 
\setlength{\figrulesep}{0.5\textfloatsep} 

\newcommand{\topfigrule}{\vspace*{-1pt}%
\noindent{\color{cream}\rule[-\figrulesep]{\columnwidth}{1.5pt}} }

\newcommand{\botfigrule}{\vspace*{-2pt}%
\noindent{\color{cream}\rule[\figrulesep]{\columnwidth}{1.5pt}} }

\newcommand{\dblfigrule}{\vspace*{-1pt}%
\noindent{\color{cream}\rule[-\figrulesep]{\textwidth}{1.5pt}} }

\makeatother

\twocolumn[
  \begin{@twocolumnfalse}
\vspace{3cm}
\sffamily

 \noindent\LARGE{\textbf{Structured randomness: Jamming of soft discs and pins}} \\
  \vspace{0.3cm} \\

\noindent\large{Prairie Wentworth-Nice,\textit{$^{a\ddag}$} Sean A. Ridout,\textit{$^{b}$} Brian Jenike,\textit{$^{a}$} Ari Liloia \textit{$^{a}$} and Amy L. Graves$^{\ast}$\textit{$^{a}$}} \\

 \noindent\normalsize {Simulations are used to find the zero temperature jamming threshold, $\phi_j$, for soft, bidisperse disks in the presence of small fixed particles, or ``pins",  arranged in a lattice. The presence of pins leads, as one expects, to a decrease in $\phi_j$.  Structural properties of the system near the jamming threshold are calculated as a function of the pin density.   While the correlation length exponent remains  $\nu = 1/2$ at low pin densities,  the system is mechanically stable with more bonds, yet fewer contacts than the Maxwell criterion implies in the absence of pins.  In addition, as pin density increases,  novel bond orientational order and long-range spatial order appear, which are correlated with the square symmetry of the pin lattice.}
\\


 \end{@twocolumnfalse} \vspace{0.6cm}

  ]

\renewcommand*\rmdefault{bch}\normalfont\upshape
\rmfamily
\section*{}
\vspace{-1cm}


\footnotetext{\textit{$^{a}$~Dept. of Physics and Astronomy, Swarthmore College, Swarthmore, PA 19081, USA. Fax: 01 610 328 7895; Tel: 01 610 328 8257; E-mail: abug1@swarthmore.edu}}
\footnotetext{\textit{$^{b}$~Dept. of Physics and Astronomy, University of Pennsylvania, Philadelphia, PA 19104, USA. }}

\footnotetext{\textit{$\ast$~Corresponding author}}

\footnotetext{\textit {\ddag~Present address: Dept. of Mathematics, Cornell University, Ithaca, NY 14853, USA.}}


\section{Introduction}
Over two decades ago, it was first proposed \cite{Durian1995,Cates1998, Liu&Nagel1998} that soft and granular materials can have a jammed, solid phase, which forms at sufficiently high packing fraction or pressure, and sufficiently low shear and temperature.  Now, much is known about materials in the vicinity of the zero-temperature jamming threshold, ``Point J"; not only for simple models like soft or hard repulsive, frictionless spheres  \cite{Epitome, OlssenTeitel2007, Liu&Nagel2010, Siemens&vanHecke2010}, but for particles which are non-spherical  \cite{Donev2007, Mailman2009, VanderWerf2018, Brito2018, Bester2019}, have rough and/or frictional surfaces  \cite{Silbert2010, Behringer2015}, are confined within various wall geometries\cite{DesmondWeeks2009, AshwinBowles2009, TorquatoStillinger2010}  and even active matter\cite{Henkes2011, Bi2015, Xu2018} (including work on active matter in the presence of fixed obstacles\cite{ReichhardtsActiveMatter2014}).  For frictionless soft spheres, a mixed first-second order phase transition, with upper critical dimension of $d=2$ occurs \cite{vanHecke2010, Liu&Nagel2010, GoodrichPRL2012, Goodrich2014}  at a maximally random close packing  fraction (MRP). Near Point J, there are diverging length scales \cite{Wyart2005, Goodrich2013, Shoenholz2013, ReichhardtReview,Hexner2018} and universal critical exponents  for quantities like contact number, static and dynamical length scales, characteristic phonon frequency, and shear viscosity below the transition \cite{Silbert2005, WyartWitten2005, Keys2007, vanHecke2010,Liu&Nagel2010, OlssenTeitel2011} - while exponents for elastic moduli depend on microscopic details like inter-particle potential  \cite{Epitome, vanHecke2010}. (A scaling relation for elastic energy has been shown to unify our understanding of the various critical exponents above the jamming transition\cite{Goodrich2016}.)   At zero temperature and shear, the critical density $\phi_j$ represents a state of marginal stability, where according to Maxwell's counting argument, the number of inter-particle contacts equals the number of unconstrained degrees of freedom.  For $d=2$, this situation of isostaticity, given translational invariance and a positive bulk modulus,  implies\cite{GoodrichPRL2012} that the average number of  contacts experienced by one of $N$ particles is  $Z_c = 4 - 2/N$. 

How then, will this maximally random, marginally stable structure be altered if particles are, in part, supported by elements internal to the system?  For the case of quenched disorder via randomly-placed attractive sites, it was proposed\cite{Reichhardts2012} that disorder constitutes a fourth axis of the phase diagram ... altering the position of $\phi_j$, as well as introducing a new critical threshold $\phi_p$  for the pinning of the flow of particles under an applied force.  (Because flow has been shown to be impeded both by attractive\cite{Reichhardts2012} and repulsive obstacles\cite{Nguyen2017, Peter2018, StoopTierno}, there is some leeway as to whether one uses the term "pinned" or "clogged" to describe the state of arrested particle flow, with its distinctive heterogeneous geometry and time lag over  which the flow comes to a halt.  For both types of obstacle, the state of arrested flow becomes a true jammed state in the limit of large packing fractions\cite{Peter2018}.)  Two different protocols for freezing particle positions as the jammed solid forms have been studied  in Ref. \cite{Brito2013} .  One protocol produces over-coordinated systems, a consequence being qualitative changes in the characteristic frequency of soft modes, linked to the length scale above which a system is hyperstatic\cite{Wyart2005}.   Applications to separation and sorting benefit from understanding flow in the presence of pinning sites\cite{ReichhardtOlsen2002}.  Moreover,  periodic arrangements of pins add an element of symmetry which can lead to predictable trends in kinematics studies, like the reduction of friction that accompanies kink propagation in driven colloidal solids\cite{Bohlein2012} or directional locking in clogging simulations\cite{ReichhardtsDirLocking2012, Nguyen2017}.  Here, as in those references, we study periodically placed pins.  Our pins are diminutive and softly repulsive, serving only to exclude volume.  

Our goal in this paper is to identify the jamming transition and explore structural features of the jammed solid as a function of the ratio of the number of pins to particles,  $n_f$.  In Section 3.1, we find the $T=0$ jamming probability, $P(\phi)$ vs. $\phi$, for various values of  $n_f$.  From that probability distribution function, we calculate a jamming threshold, $\phi_j(n_f)$, which is a decreasing function of $n_f$. Once unsupported particles (``rattlers") are removed, $\phi_j(n_f)$ decreases even more dramatically.  A comparison with randomly distributed pins, suggests that having an appreciable ratio of particle size to pin separation is important in determining whether the geometry of the pin lattice affects the detailed shape of $\phi_j(n_f)$.    In Section 3.2 we calculate structural quantities, beginning with the probability distribution of bond angles, $P(\theta)$. There is a transition from a uniform distribution to one with fourfold symmetry, again occurring when the ratio of particle size to pin separation is appreciable.  A numerical solution seems a necessary evil, in order to predict the intricately detailed substructure in $P(\theta)$ which arises from the bidispersity of particle sizes, packing amid pins.  We then discuss the pair correlation function $g(r)$  and scattering function $S(\vec{k})$ with $\hat{k}$ referred to lattice symmetry axes. Both bond angles and pair correlations reveal how pins impose order, whose character and magnitude can be tuned via the density of pins. In Section 3.3, we calculate contact statistics - concluding that pins support a jammed system with fewer contacts between particles, and fewer contacts overall.  We might think of jamming in the presence of a square lattice of pins as not only possessing  a modicum of order, but as ``parsemonious", requiring fewer particles, each one requiring fewer contacts, in order to form a rigid solid.

\section{Methods}
$M$ pins are placed on a square lattice in a 1x1 simulation cell with  periodic boundary conditions. $N = N_{l} + N_{s}$ large and small particles are introduced at random locations, using a uniform random number generator.  The initial packing fraction is thus  $\phi=N_{s}\pi r_{s}^2+N_{l}\pi r_{l}^2$ with $N_s = N_l$ and $r_{l}/r_{s} = 1.4$, a fraction of large-to-small particles and size ratio that previous studies have found optimal in two dimensions, to inhibit the formation of a regular, hexagonal lattice\cite{ReichhardtReview}.    Pin radii $r_{pin}$  are much smaller than those of particles; $r_{pin} =  r_{s}/1000$, and variation around this value had a negligible effect on results. 
 All particles in our system are soft disks, with a short range harmonic interaction potential given by
\begin{align} V = \left\{
\begin{matrix}
0 & r_{ij}>d_{i,j}\\
\frac{1}{2}\epsilon(1-\frac{r_{ij}}{d_{ij}})^2 & \mbox{otherwise}
\end{matrix} 
\right. ,
\end{align}
where $r_{ij}$ is the distance between the centers of particles $i$ and $j$, $d_{ij}$ is the sum of the radii of the two particles, and $\epsilon \equiv 1$ for this zero-temperature study. 
It is known that the probability distribution of jamming thresholds is protocol-dependent \cite{Chaudhuri2010, Bertrand2016}.  Structural and scaling properties may be noticably different if the protocol yields a jammed configuration which is hyperstatic \cite{Schreck2011};  and even if isostatic, if the starting configuration is a hard-particle liquid exceeding a certain packing fraction\cite{Ozawa2012}.   We adopt a simple, athermal protocol which, in the absence of pins, is expected to yield a jamming transition at the lowest limit of any ``line" of jamming thresholds\cite{Xu2018}, with the canonical structural and scaling properties described in O'Hern et al. \cite{Epitome}.   The energy of a random configuration of soft discs is minimized via a conjugate gradient algorithm\cite{conjugate_gradient}, after which configurations are tested for mechanical stability, and unsupported particles (``rattlers") are removed.  1000 random seeds are used to generate initial particle configurations for each  $[M, N]$ pair.  Packing fractions $\phi$ are chosen to completely span the jamming transition for a given value of $M$. 
  
\begin{figure}[h]
\centering
\begin{subfigure}{.25\textwidth}
  \centering
  \includegraphics[width=1.18 \linewidth]{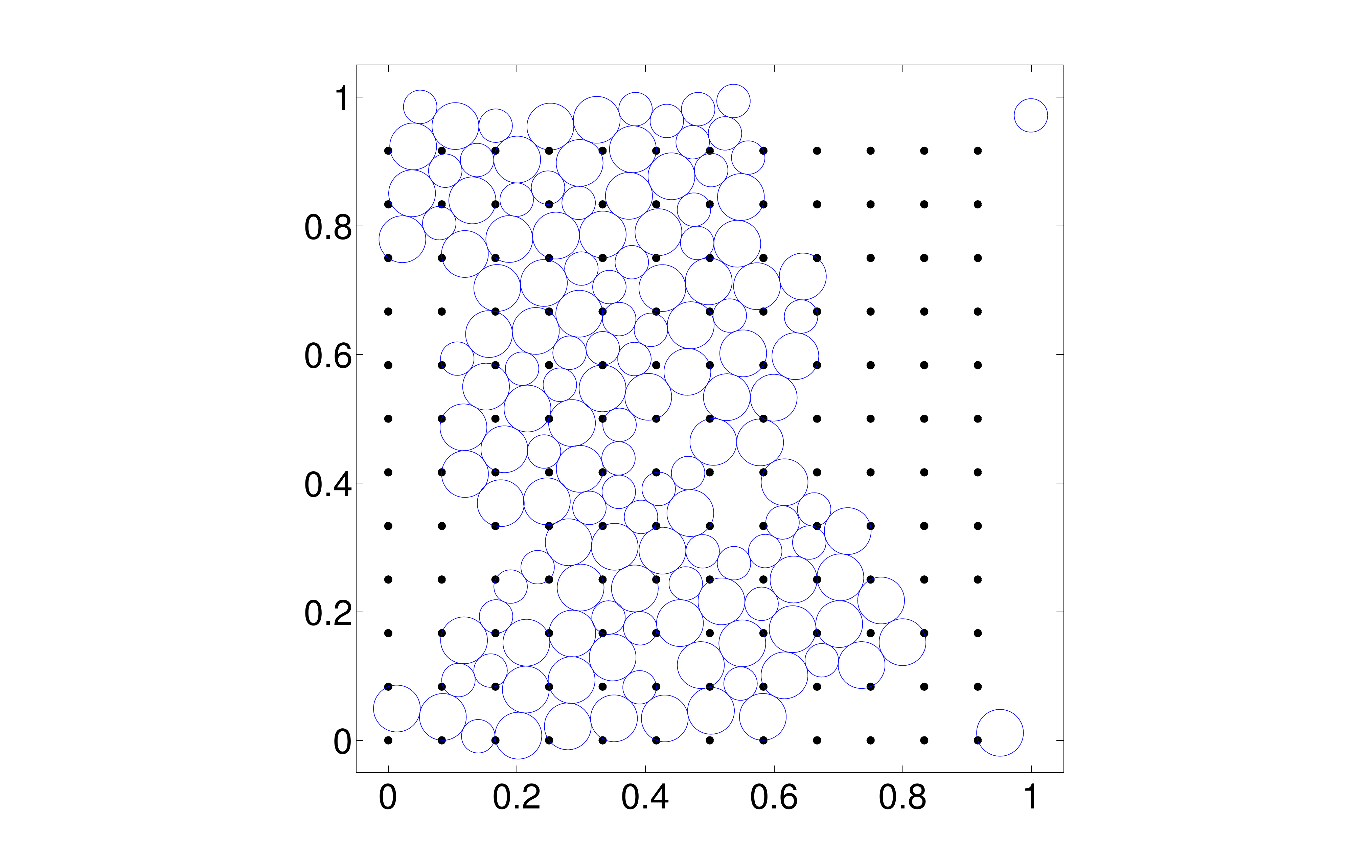}
  \caption{}
  \label{JamNoPerc}
\end{subfigure}%
\begin{subfigure}{.25\textwidth}
  \includegraphics[width=1.18 \linewidth]{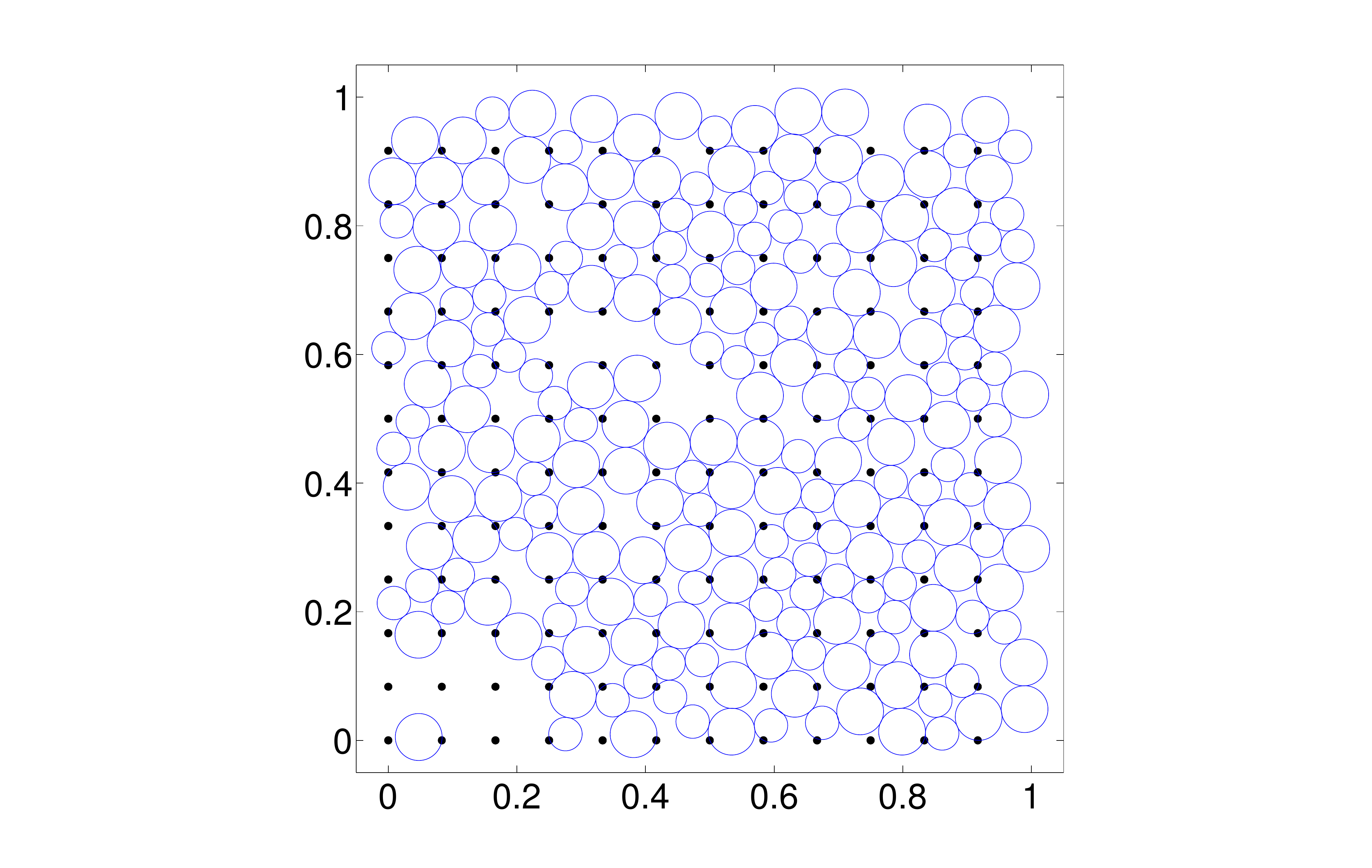}
  \caption{}
  \label{JamAndPerc}
\end{subfigure}
\caption{Two final configurations, both initialized with $N=249$ particles and $M=144$ pins. The pins are enlarged for visibility. (a) Configuration jams, but does not percolate. (b) Configuration jams and percolates. }
\label{PercConfigs}
\end{figure}

 The simulation halts when a chosen tolerance for changes in the gradient of the energy is reached.  Near the jamming threshold, this energy is typically less than $10^{-6}$ of the energy scale for a single pair interaction.  For the data discussed below, we not only assert that particles in a  jammed  solid are mechanically stable, but we ask that the system ``percolates", so that there is a connected path between box top and bottom sides, and between left and right sides.  This criterion is not needed in the absence of pins, when mechanical stability occurs only if the system percolates. It is, however, a needed criterion in systems  , in which a short-ranged attraction between jammed particles permits the creation of non-spanning rigid clusters~\cite{Lois2008, Koeze2020}.    Figs. \ref{JamNoPerc}, \ref{JamAndPerc} illustrate two  final configurations in a system with $M = 144$,  hence $a = 0.0833$. This is sufficiently close to the particle size ($r_{s} = 0.0260382$)  that, at jamming, equilibration may result in situations like Fig. \ref{JamNoPerc}, where no cluster of non-rattlers spans the system from left to right.  At the highest pin densitiy discussed below, approximately $1/4$ of jammed configurations fail to percolate.  Non-percolating equilibrated configurations like Fig. \ref{JamNoPerc} were excluded from further analysis.

In our work, two protocols were examined, and the second was adopted for results shown in Fig. ~\ref{FSS} and beyond.  At issue is the state space controlled by three parameters: $N$, $\phi$, and $M$ which cannot be neatly collapsed. Suppose one utilizes a simulation cell of fixed size, with a fixed number of particles, $N$ and pins $M$.  In this "first protocol",  $\phi$ is varied by changing $r_s$.  This may be repeated for larger values of $N$, and finite-size scaling (FSS) arguments applied in order to extrapolate to the $N \rightarrow \infty$ limit.   However, the pin lattice introduces an additional length scale, $a = 1/\sqrt{M}$.   The protocol just described involves changing the ratio $r_s/a$ as one scans through $\phi$ values in order to locate $\phi_j$.  If one suspects that the ratio of particle size to pin separation is a meaningful control parameter,  one seeks a different approach.  Thus, we employed a ``second protocol":  $r_{s}$ is fixed and $N$ is varied in order to vary $\phi$ in the neighborhood of $\phi_j$.  The range of $N$ values does not need to be large (less than $10\%$ for any value of $M$ in the current study) in order to span the phase transition. 
 
 With either protocol (rather than one where the system is relaxed from an initially overjammed configuration)  sometimes the minimization procedure is not precise enough to identify a very slightly unstable configuration as such.  Thus, we discard a very small fraction (less than $1\%$) of configurations with an insufficient number of bonds, $N_{iso}$,  required to satisfy the number of degrees of freedom. These must necessarily have zero vibrational frequency modes, and be unstable.  In packings without pins, we have checked that this criterion perfectly identifies a group of packings which also have unusually high energies and zero (or negative) bulk moduli.
    
\section{Results}
\subsection{Jamming probability}
We examine the jamming transition by calculating $P_j(\phi)$,  the probability of jamming as a function $\phi$  for different system sizes $N$, and different numbers of pins, $M$.  
A sigmoidal fitting form which estimates the center, $\phi_j$, and width, $w$ is: 
\begin{equation}
P_j(\phi)=\frac{1}{(1+b \ e^{(\phi_j - \phi)/w})^{1/b}}.
\label{Sigmoid}
\end{equation}
Eq. ~\ref{Sigmoid} is a Richards sigmoid which, for $b = 1$ is an isotropic, logistic sigmoid. This 2-parameter logistic sigmoid is used for  $M = 0, \ 36, \ 81$ and $100$ .  For $M=144, \ 169$,  the deviation from $b=1$ is significant in order to account for anisotropy about $\phi_j$; and thus the 3-parameter form of Eq. ~\ref{Sigmoid} is utilized. 

\begin{figure}[h!]
	\centering
  \includegraphics[width=1.0\linewidth]{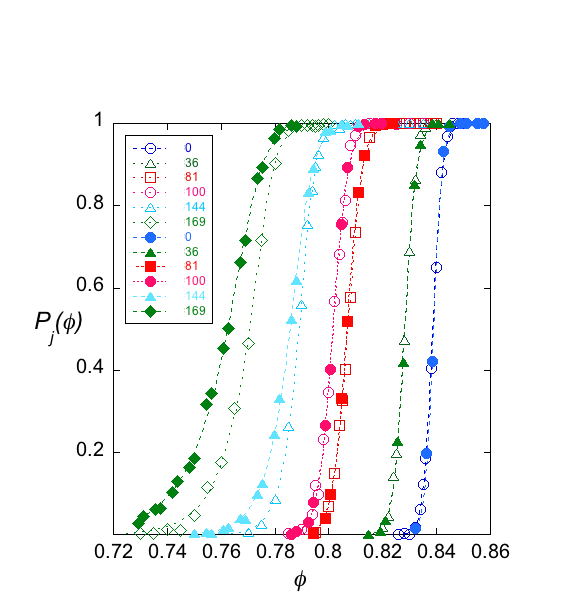}
  \caption{$P_j(\phi)$ for systems with number of pins, $M$, given in legend and $N = 256$ particles (first protocol, open symbols) and $N \in [231, 261]$ particles (second protocol, closed symbols). Dashed lines indicate sigmoidal fits. }
  \label{PjN256}
\end{figure}
 Fig. \ref{PjN256} and Table \ref{phijTable} illustrate that the two protocols yield consistent results for $P_j$  in the limit of small  $M$ . However, there is a difference in the two protocols for the densest pin lattices.  As $M$ becomes comparable to  $N$, the second protocol (which preserves the ratio of  $a$ to the particle size) yields a lower $\phi_j$, and also yields a transition which becomes steadily wider as $M$ increases.  Unless otherwise noted,  data discussed henceforth were calculated with the second protocol. 

\begin{table}[h!]
\centering
  \begin{tabular*}{0.48\textwidth}{@{\extracolsep{\fill}}lll}
  \hline
 Number of pins & Second Protocol $\phi_j$, $w$ & First Protocol $\phi_j$, $w$ \\
  \hline
$M=0$ & $0.838 \pm 1$ , $0.0018 \pm 2$ & $0.838 \pm 1$ , $0.0020 \pm 2$ \\
\hline
$M=36$ & $0.828  \pm 1 $, $0.0022 \pm 2 $ & $0.828  \pm 1 $, $0.0022 \pm 2$ \\
 \hline
 $M=64$ & $0.820  \pm 1 $, $0.0022 \pm 2 $ & $0.8208  \pm 1 $, $0.0022 \pm 2$ \\
 \hline
 $M=81$ & $0.807  \pm 1 $, $0.0027 \pm 2$ & $0.8078  \pm 1$, $0.0028 \pm 2$  \\
 \hline
 $M=100$ & $0.802  \pm 1$, $0.0028 \pm 2$ & $0.802 \pm1 $ ,  $0.0028\pm 2$ \\
 \hline
 $M=144$ & $0.787\pm 1$, $0.0036 \pm 3$ & $0.790 \pm 1$,  $0.0025 \pm 3$ \\
 \hline
 $M=169$ & $0.766 \pm 2$, $0.0046 \pm 4 $  & $0.773\pm  2$,  $0.0033 \pm 3$ \\
\end{tabular*}
\caption{Calculated critical point $\phi_j$, and width $w$ of transition, using sigmoidal fit to 1000 realizations per $\phi$ value.  Systems contained $M$ pins and $N\approx 256$ particles before equilibration. }
\label{phijTable}
\end{table}

\begin{figure}[h]
\centering
\begin{subfigure}{.24\textwidth}
  \centering
  \includegraphics[width=1.07 \linewidth]{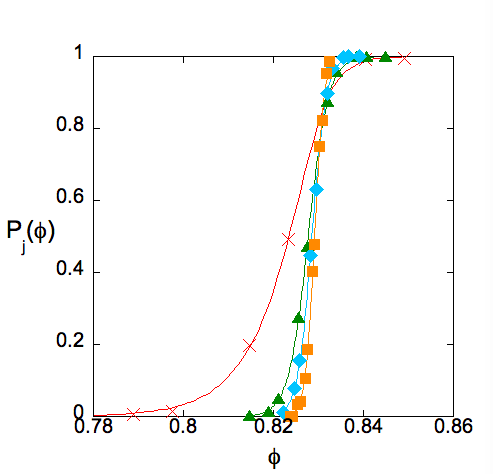}
  \caption{}
  \label{FSSnewpreScale}
\end{subfigure}%
\begin{subfigure}{.24\textwidth}
  \includegraphics[width=1.15 \linewidth]{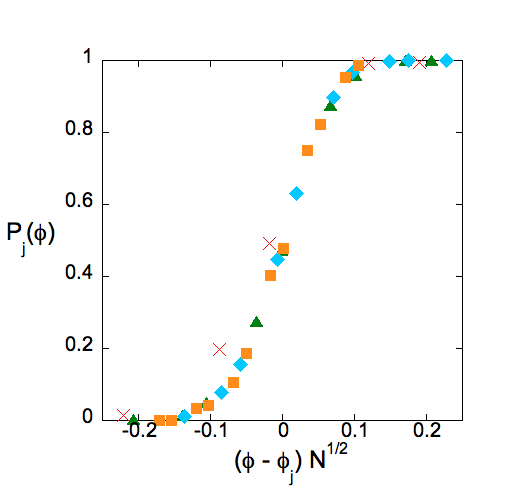}
  \caption{}
  \label{FSSnew}
\end{subfigure}
\caption{Jamming at a pin density of  $n_f = 0.14$ for four systems of different size regimes, with $N \approx 64, \ M = 9$ (red crosses), $N \approx 256, \ M= 36$ (green triangles), $N\approx 455, \ M = 64$ (aqua diamonds) and $N \approx 1024, \ M= 144$ (orange squares).  (a)  Probability of jamming as a function of packing fraction. Richard's sigmoid fit: $N \approx 64$,   logistic sigmoid fit: $N \approx 256, \ 455, \ 1024$. (b) FSS version of (a), illustrating collapse of $P_j$ data for three sufficiently large $N$ values. }
\label{FSS}
\end{figure}

 As with other phase transitions,  FSS allows one to approach the limit $N \rightarrow \infty$ systematically  to infer critical properties. Figs.  \ref{FSSnewpreScale} and  \ref{FSSnew} show that for the same value of $n_f$, $P(\phi)$ for different $N$ can be collapsed onto one scaling curve, just as would be true without pins\cite{GoodrichPRL2012, Goodrich2014, Graves2016}  \footnote{Scaling behavior with $n_f$ has also been proposed in the limit of dilute, fixed particles\cite{Graves2016}.}.    
Here, one plots $P_j$ as a function of the rescaled distance to the critical point: $(\phi-\phi_j(N, n_f)) \ N^{1/2}$.   (Though $N$ values vary slightly with this protocol, $n_f \approx 0.14$ is constant to two significant figures for the data shown.)  The rescaling exponent \ $1/2$ is determined from the width at half maximum of the distribution of jamming thresholds \cite{Epitome}.  At the smallest size $N=64$ it is not surprising that this procedure, sans corrections to scaling, will fail~\cite{Epitome, Vagberg2011}. 

\begin{figure}[h]
\centering
  \includegraphics[width=1.0 \linewidth]{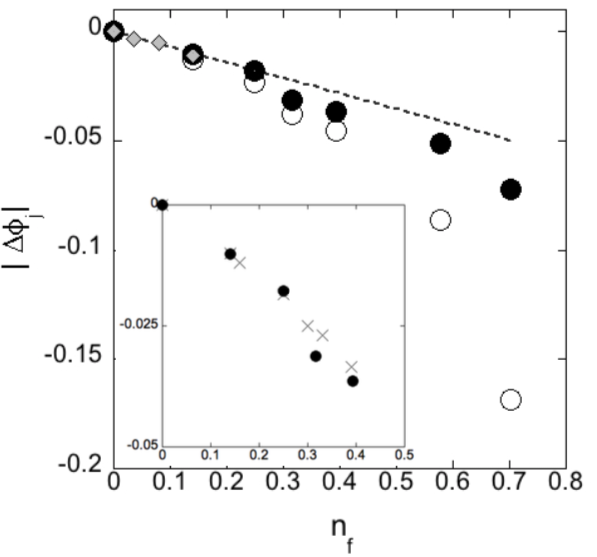}
  \caption{Deviation of jamming threshold from the zero-pin value, as a function of pin density $n_f$.  Filled circles: Threshold in terms of initial volume fraction, for $N \approx 256$; grey diamonds: $N = 1024$. Open circles: Final volume fraction, once rattlers are removed, for $N \approx 256$.  Error bars are smaller than data points.  Linear fit with slope $-0.071$ is shown, as is consistent with data for $n_f \leq 0.25$.  Inset: Filled circles: Jamming threshold as in main figure, for pins configured in square lattice. Crosses: Jamming threshold for random distribution of pins. }
  \label{phijvsnf}
\end{figure}

Studies of jamming in which a fraction particles are immobilized have been unanimous in observing a reduction in the jamming threshold.  While in previous studies obstacles were slightly larger\cite{StoopTierno} or on par with the size\cite{Brito2013, ReichhardtReview, Nguyen2017,Peter2018} of particles, our pins make a negligible contribution to the volume fraction. Perhaps this makes the 
the decrease in $\phi_j$ in Fig. ~\ref{phijvsnf}  less surprising than in previous studies, for in the limit of low pin densities, pins support rigidity without adding volume. Nevertheless,  at higher densities,  the geometry of the pin lattice may not only to generate order (see Section 3.2) but adjacent pin proximity may interfere with the formation of a jammed solid\footnote{Preliminary data indicate that,  the rule of monotonic decrease seen in Fig. 4 can be violated.  These results,  which include additional pin geometries, will appear in a forthcoming publication.}. 

The inset to Fig. ~\ref{phijvsnf} 
provides a comparison between the square pin lattice and an additional geometry: pins placed at random positions in the simulation cell, which has appeared in both experimental work on paramagentic colloids being driven around silica obstacles\cite{StoopTierno} and simulations\cite{Brito2013, ReichhardtReview, Peter2018}.  As one might expect, the inset shows agreement between the two geometries in the dilute pin limit.

It has been argued \cite{Reichhardts2012, Graves2016} that jamming occurs when the inter-pin separation is equal to the correlation length, $\xi \sim (\phi-\phi_j)^{-\nu}$.   Thus one would predict that in $d$ dimensions, pins will lower the jamming threshold by the amount:
\begin{equation}
\Delta \phi_j (n_f) \equiv \phi_j (n_f) - \phi_j(0) \propto n_f^{ \ \ 1/d\nu}
\label{correlation}
\end{equation}
Here $d=2$, so a linear fit of  $\Delta \phi_j (n_f)$ implies that $\nu=\frac{1}{2}$, the accepted value for spherical, repulsive particles in the absence of friction\cite{Durian1995, Siemens&vanHecke2010}.  Fig. \ref{phijvsnf}  suggests that for $n_f \leq 0.25$, the data is approximately linear; a power law fit yields $|\Delta \phi_j (n_f)| \propto n_f^{\ \alpha}$  with $\alpha = 0.91$  and $\alpha = 0.95 \pm 0.10$ for $N = 256, \ 1024$ respectively.  If we assume that $\alpha = 1$ (linear fit) the slope is 
$\frac{\partial \phi_j}{\partial n_f} = -0.071,  -0.072  \pm 0.002 \ $ for  $ N = 256, \ 1024 $ respectively.  The slope value is seen to depend on the details of the interaction between particles and obstacles\cite{Reichhardts2012, Graves2016, Peter2018}.  More mportantly, deviation of the data in Fig. \ref{phijvsnf}  from linearity at higher $n_f$ suggest that sufficiently dense, ordered pins do more than single out the correlation length of a highly random packing.  

The thresholds for square and random lattices both deviate from linearity and begin to disagree with one another occurs at approximately $n_f = 0.28$.  The answer to "Why this density?" seems to incorporate the ratio of pin size to typical inter-pin spacing, $r_s/a$:
\begin{equation}
r_s/a =  \sqrt{\pi  \ \frac{1+ 1.4^2}{2} \  \phi \ n_f \ \ }.
\label{lambda}
\end{equation}
Using Eq. ~\ref{lambda} and fitting to $\phi_j(M)$ from the square lattice data in Fig. ~\ref{phijvsnf} yields 
\begin{equation}
r_s/a \ = \ 0.39 \ n_f^ {\ 0.46}.
\label{lambdaToo}
\end{equation}
 Values of $\phi_j$ for square and random pin configurations begin to disagree at approximately $r_s \ = \ 0.22 a$; when small and large particles have diameters which are 40\% and 60\% of the typical pin separation.  That the relevant length scale when lattice identity affects jamming is a microscopic length scale, on par with particle size, also applies to the data on bond structure in Section 3.2.  It agrees with the size scale - a couple of particle diameters - on which confining boundaries produce inefficient packing, featuring a lower packing fraction and some evidence of square-like packing, in a study of hard discs confined by walls ~\cite{DesmondWeeks2013}.

Since $\phi_j$ is traditionally pitched as a criticial initial packing fraction, one may also ask about the final packing fraction after removal of particles which do not alter the mechanical stability (or percolation) of the system.   The open circles in Fig. ~\ref{phijvsnf} show the final volume fraction, after one eliminates these rattlers.  As $M$ increases, so does the volume of rattlers at $\phi_j $, leading to a dramatically smaller packing fraction of particles needed for rigidity.   A technological benefit of creating a jammed solid supported by pins would be a less dense material.  Below, we show that there is an onset of both local (bond) and global (positional) ordering as the pin lattice density increases. This suggests an additional benefit: the ability of pins to induce elastic and transport properties which are anisotropic and adjustable. 

\subsection{Angular and spatial ordering}
\begin{figure}[h]
\centering
  \includegraphics[width=.8 \linewidth]{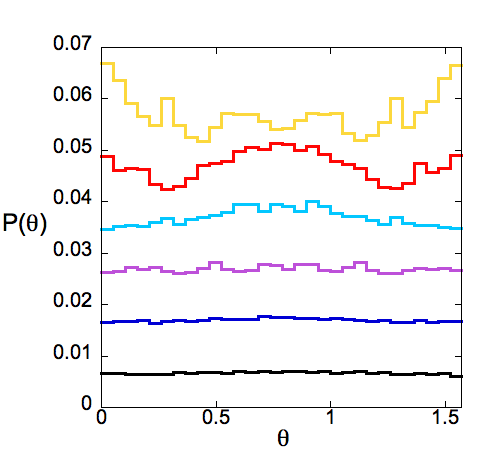}
  \caption{Histograms proportional to the probability $P(\theta)$,  that a bond makes angle $\theta$ with the x-axis.  Each data set corresponds to the system parameters closest to the fraction at jamming $\phi_j$  from sigmoidal fits, and given Table \ref{phijTable}).  Different numbers of pins $M$ corresponding to different size ratios at jamming $r_s/a$ are shown in black: $M=0, \ r_s/a = 0$;  dark blue: $M=36, \  r_s/a =  0.158 $; purple: $M=81, \ r_s/a= 0.234$;  light blue: $M=100, \ r_s/a = 0.260$; red: $M=144, \ r_s/a =  0.312$; and yellow: $M=169, \  r_s/a = 0.338$. Histograms for successive parameter values are displaced from each other vertically by $0.01$ for ease of viewing. }
  \label{Ptheta}
\end{figure}

Two particles whose radii overlap can be thought of as sharing a ``bond",  oriented in the direction between the particle centers. The geometry of the bond network is central to the theory of jamming; controlling the fragility of the jammed state,  elastic properties,  the phonon spectrum, scaling exponents, and evolving in response to compression and shear\cite{Cates1998, MajumdarBehringer2005, Zhang2010}. The angle formed by a pair of bonds which carry the largest forces reveals a distribution which supports the picture of ``force chains" \cite{ZhouDinsmore2009, DesmondWeeks2013} .  In the presence of pins, we must first ask a more basic question about individual bonds: Are these isotropically distributed in angle space?  To capitalize on the comparison of square and random pin lattice, we will answer this question in terms of the variable  $r_s/a$; which can be mapped back to $n_f$ at the jamming threshold via Eq. ~\ref{lambdaToo}. 
Fig. \ref{Ptheta}  compares the distribution of bond angles with  $N \approx 256$ systems for various pin densities.   One sees the distribution become progressively more anisotropic as $M$ increases.  The distribution of bonds $P(\theta)$ has fourfold symmetry as one expects given a square pin lattice.  (Fig. \ref{Ptheta} only shows angles in the range $\theta \in [ 0, \pi/2 ]$; bonds with $\theta \in [ \pi/2, \pi ] $ are averaged with the data shown.) The reflection symmetry about  $\theta = \pi/4$ is clear.   

\begin{figure}[h]
\centering
  \includegraphics[width=.8 \linewidth]{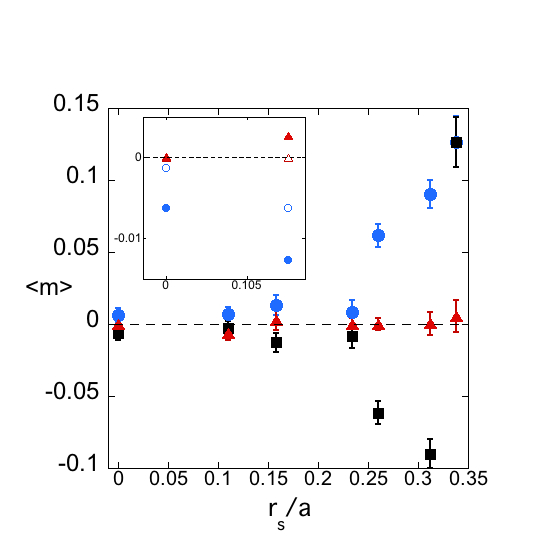}
  \caption{Order parameter as a function of particle-lattice size ratio.  Black squares: $Re <m>$,  red triangles: $Im <m>$, and blue circles: $ |<m>|$.   Inset:  Real and imaginary parts of $<m>$ showing both  $N=256$ (solid symbols) and $N=1024$ (open symbols) data. }  
  \label{OrderParameter}
\end{figure}

Fig. \ref{OrderParameter}  shows the order parameter $ < m > \equiv < e^{i 4 \theta} > $  for $N \approx 256$ systems.   The magnitude of the real part of this order parameter, given the choice of axes, is approximately equal to  $|<m>|$. Apparently,  an angular ordering transition occurs somewhere between $ r_s / a = 0.23$ and $0.26$; in good agreement with the density at which the the threshold, in the previous section, shows evidence of being affected by the pin lattice geometry.   While $|<m>|$ continues to increase above the transition, note the change in sign of $Re <m>$ at the densest lattice studied (yellow curve in Fig. \ref{Ptheta})  which correlates with the change in the most-probable bond angle  from $\theta = \pi / 4$ for less dense lattices, to $\theta = 0, \ \pi$. While it appears that $ < m > $  is nonzero for all values of $M > 0$ studied, this may be a finite-size effect having to do with the square simulation cell with periodic boundary conditions, which creates a slight amount of ordering even for $M=0$\cite{DesmondWeeks2013}.  As a check on this, the inset of the figure shows both $N=256$ and $1024$ data for $r_s / a = 0, \ 0.167$.    For the larger system (open symbols) the order parameter is significantly closer to zero.   The T=0 jamming transition constitutes an out-of-equilibrium,  phase transition.    Nevertheless, the abrupt increase in $|<m>|$ with cubatic ordering from pins is reminiscent of a  phase transition, such as the isotropic to nematic transition in uniaxial liquid crystals in the presence of an orienting field \cite{Singh}.  Statistical ensemble ideas have successfully described jamming \cite{BiDaniels2015}  and perhaps will allow one to map out a bond ordering transition using pin density as a control parameter. 

\begin{figure}[h]
\centering
  \includegraphics[width=1.0\linewidth]{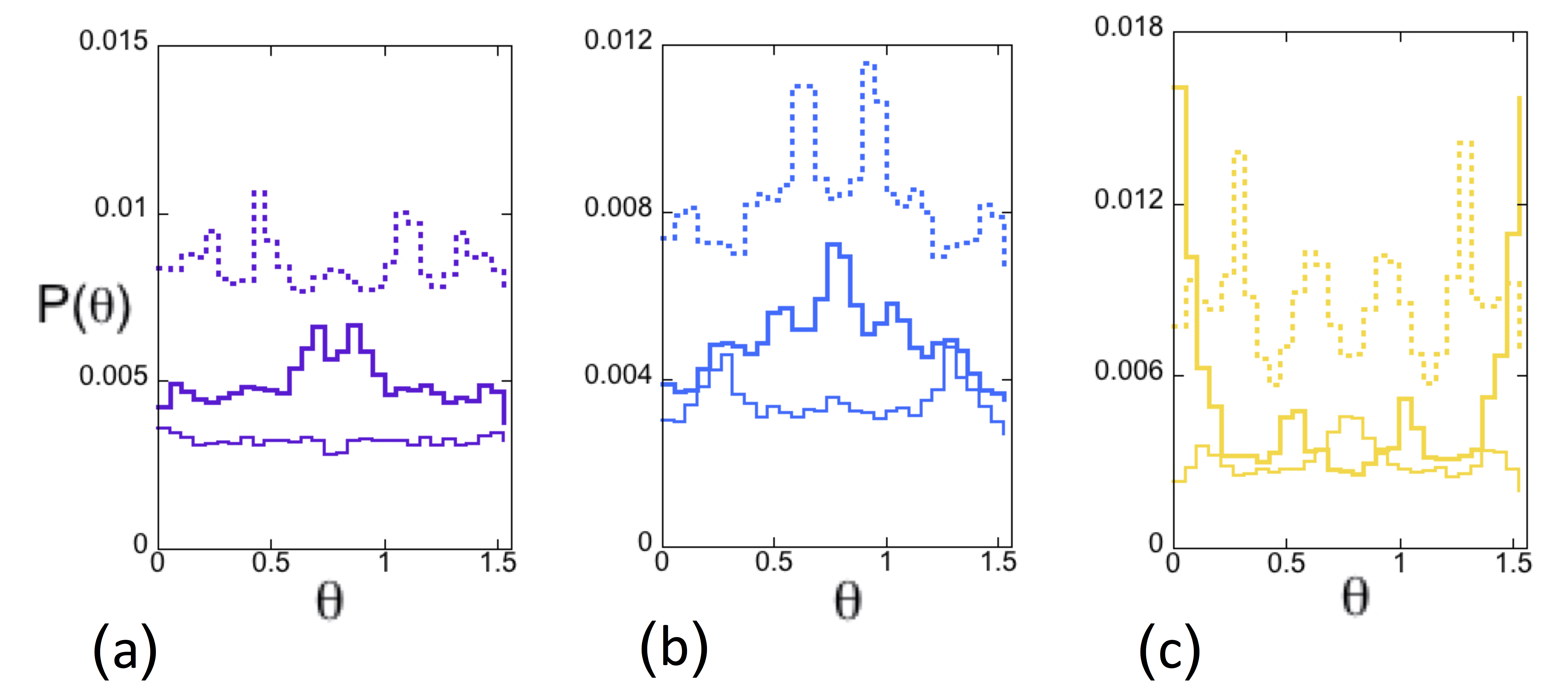}
  \caption{$P(\theta)$ for different types of bond. Thick lines: large-large, thin: small-small, dotted: large-small. Colors indicating $M$ value are as in previous figures. a) purple: $81$; b) light blue: $100$; c) yellow: $169$.}
  \label{PthetaParticleSizes}
\end{figure}

A bidisperse distribution results in particle-size-dependent preferences for certain bond angles. Thus, detailed features of $P(\theta)$  in Fig. \ref{Ptheta} can be traced to bonds between particles of specific sizes, as exemplified in Fig. \ref{PthetaParticleSizes}. 
The large-large bonds tend to be oriented near $\theta = \pi/4$ for $M=81, 100$; however a couple of other favorable orientations appear as side peaks in Figs. ~\ref{PthetaParticleSizes}a,b.  At the highest pin density studied, $M=169$ of Fig.~\ref{PthetaParticleSizes}c, large-large bond probabilities have a peak at approximately $\theta=\pi/6, 2 \pi/3$ (also true for $M=100$) but the most probable orientation is horizontal or vertical.  In contrast, at this pin density, the small-small bond angles have a preferred orientation of $\theta = \pi/4$. Small-small bonds show little orientational structure at $M=81$,  and at $M=100$ are most likely to be oriented at $ \theta =\pi/12, \  5 \pi/12$; coinciding with one set of large-large bond peaks.  Large-small bond probabilities have, for $M=81$  a small local maximum at $\theta = \pi/4$. But it is overshadowed by peaks at other angles, present for all three $M$ values in Fig.~\ref{PthetaParticleSizes}.  The take home message is that the details of the bond angle distribution depend in an intricate way on particle sizes $r_s, \ r_l$ and pin separation, $a$.   It is worth noting that there is only a slight degree of particle size segregation;  whether same-sized or differently-sized particles are more likely to share a contact varies only slightly with $M$.  Segregation is largest at $M = 169$, where differently-sized particle contacts exceed same-sized ones by $4\%$. 

\begin{figure}[h]
\centering
\begin{subfigure}{.24\textwidth}
  \centering
  \includegraphics[width=1.0\linewidth]{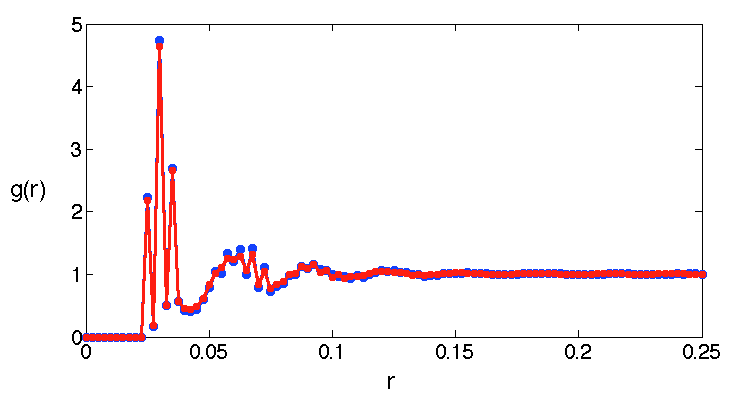}
  \caption{}
  \label{gofr1024short}
\end{subfigure}%
\begin{subfigure}{.24\textwidth}
  \includegraphics[width=1.0 \linewidth]{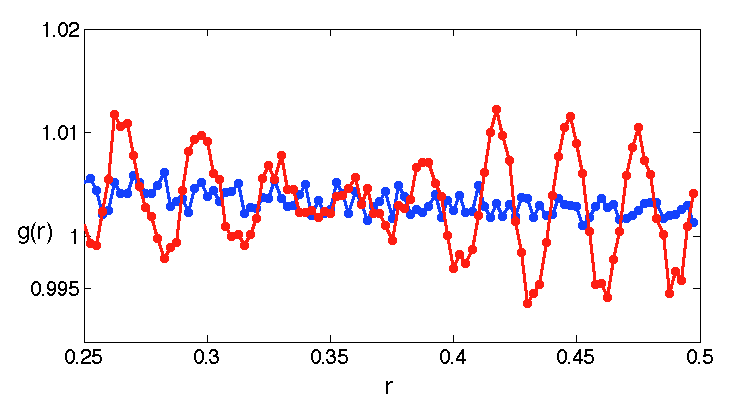}
  \caption{}
  \label{gofr1024long}
\end{subfigure}
\caption{Pair correlation function, $g(r)$ for $N =1024$ particles at the jamming threshold.   Blue symbols: $M=0$.  Red symbols: $M = 144, \ n_f = 0.141$.}
\label{gofr1024}
\end{figure}

The pair correlation function between particle centers, $g(r)$ with $r_s =0.0133$, and $r_l = 0.0186$ is shown
in Fig.  ~\ref{gofr1024} for $N = 1024$ particles - with zero pins (blue) and $M = 144$ pins (red), so that $a=0.0833$.  (As in the inset of Fig. ~\ref{OrderParameter}, $N=1024$ is utilized here to lessen the finite-size effects on structure, and distinguish them from the effect of pins \cite{DesmondWeeks2013}.)  At the modest pin density $n_f =0.141$,  Fig. ~\ref{gofr1024short} indicates that pins do not dramatically change $g(r)$ at distances $r$ which are within the first few ``solvation shells" of a reference particle.     Structure related to the bidisperse system is visible;  for example, the first three peaks correspond to $r  \ \approx \ 2 r_s,  \ r_s + r_l , \ $ and $2 r_l$.  However,  as seen in Fig. ~\ref{gofr1024long}, pins are responsible for the persistence of this order across the simulation cell.  These regular oscillations in $g(r)$ have a spatial period of $0.030 \pm 0.005$, the typical separation between neighboring particles.  These oscillations feature an amplitude modulation, which can be explained by the superposition of contacts at large-large, large-small, and small-small contact distances: $ 0.037, \ 0.032, \ 0.027$.   The width of the ``beat pattern" of three such superposed sinusoids is quite comparable to the value of $0.20$, seen in Fig. ~\ref{gofr1024long}. 

Structure seen in $g(r)$  carries over into its Fourier transform, $S({k})$. Moreover,  $S(\vec{k})$  for vector $\vec{k}$ reveals any long range order in structurally-relevant directions.  For a perfect square lattice, these directions would simply be all integers $[h \  l]$.   We define a normalized scattering intensity as this structure factor:
\begin{align}
S(\vec{k})=\frac{1}{N}\sum_{j=1}^{N}\sum_{k=1}^{N}e^{-i\vec{k}(\vec{R_j}-\vec{R_k})}
\label{Sk}
\end{align}
where the sums in Eq. ~\ref{Sk} extend over pairs of particles.  
\begin{figure}[h]
\centering
  \includegraphics[width=1.0\linewidth]{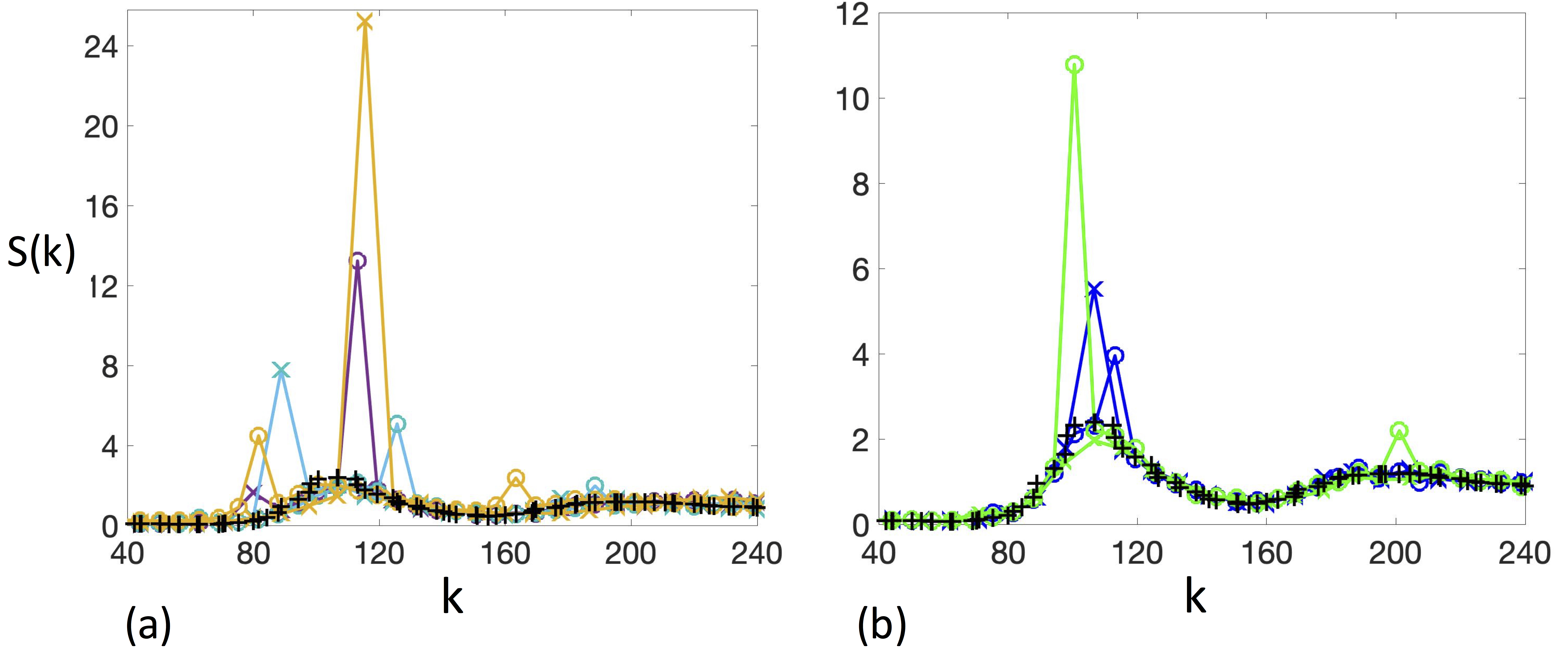}
 \caption{Particle scattering function $S(\vec{k})$ vs. wave number $k$, in units where the box is of linear size 1.  Colors indicating $M$ value are:  Black: $0$, dark blue: $36$, green: $64$, purple: $81$, light blue: $100$,  yellow: $169$.  The $+$ symbols signify $M=0$ data.  For other $M$ values, crosses signify $\hat{k}$ oriented at $45^o$ with respect to the row direction of the pin lattice; circles signify $\hat{k}$ oriented at $0^o$.   }
\label{Sofk}
\end{figure}

The structure factor $S(\vec{k}) \equiv S(k) $ for zero pins is shown with $+$ symbols in Fig. ~\ref {Sofk}, while circles in Figs.  ~\ref{Sofk}a, b depict  $S(k_{10})$ where $k_x = k, \ k_y = 0$ and crosses depict $S(k_{11})$, where $k_y= k_x = \sqrt{2} k$.  Colors indicate $M$ values.  Discretization due to the periodic boundaries of the 1x1 simulation cell restricts the resolution in k space to $ \Delta k = 2 \pi$. The horizontal scale is chosen to focus on the region relevant to the first couple of "solvation shells".  A peak from particles separated by precisely $2 r_s$, $r_s+r_l$ or $2r_l$  along each lattice direction would fall roughly at  $k = 121, \ 101$ or $86$.  The main message of Fig. ~\ref{Sofk}  is the existence of lattice-induced peaks which are correlated with the positions of pins. Were it simply the case that particles took on the crystalline symmetry of a perfect square,  peaks would fall  at the reciprocal lattice vectors  $k_x = n \Delta k, \ k_y = m \Delta k$ with $n, m$ integers. Small amounts of positional disorder and finite size effects from simulation would result in recognizable modulation of peak heights and widths \cite{Guinier}. For our finite system the peaks for $n=m=1$ and $n=1, \ m=0$ would be equal in height. With pins, $S(\vec{k})$ is neither that of an isotropic system, nor a square crystal with positional disorder.  In Fig.  ~\ref{Sofk}, discrete peaks rise from the isotropic background. The height of the tallest peak increases as the pin density increases.  

  All of the peaks in Fig. ~\ref{Sofk}a have this in common: They signify particles whose separation, projected onto the direction of $\vec{k}$,  is half of the inter-pin spacing.  This reflects the compromise which particles make to close-pack while avoiding pins with the chosen separation.   For $M=81$, $a  = 0.111$: The large peak (circle) at k = 113, has $2 \pi /k \ = a/2$. The much smaller peak (cross) at $k = 80$ implies $2 \pi /k \ = \sqrt{2} a / 2$.   For $M=100$,  $a  = 0.1$:  Again one sees two per unit cell in both horizontally and diagonally. For $M=169$,  $a  = 0.0769$:  The dominant peak (cross) at $k = 115.5$ indicates two particles per lattice unit cell projected diagonally.  

At the pin densities shown in Fig. ~\ref{Sofk}a, a pair of bidisperse particles, in contact but non-overlapping, do not in general ``fit" within an $a \times a$ square region with pins at its four vertices.  Analyzing these data presents an enormously complicated packing problem, even if only particle pairs are considered.  One speculates that such packing constraints drive the qualitative shift from ordering in the horizontal or vertical directions,  to ordering along a diagonal in the lattice.   On the other hand, the trend shown in Fig. ~\ref{PthetaParticleSizes} for bond angles is {\em opposite}, with a bond probability density maximum at $\theta = 45^o$ at lower pin densities, which has shifted to $\theta = 0$ at higher densities.  Local bond  and long-range positional order are, broadly speaking, two distinct results of the pin lattice. 

 Supporting this notion is the fact that the more dilute pin lattices are also able to promote long-range positional order. 
This is seen both in Fig. ~\ref{gofr1024} and in Fig.  ~\ref{Sofk}b, which depicts the two lowest pin-densities studied.  These have no significant bond order according to Fig. ~\ref{OrderParameter}. In Fig. ~\ref{Sofk}b for $M=36$, $a = 0.167$: The peak (circles) at $k = 113.09$ give $2\pi/k = 0.0556 = a/3$. The naïve picture is one of three particles spanning a lattice unit cell horizontally (or vertically). The peak (crosses) at $k = 106.6292$ implies $2\pi/k = 0.0589 = a \sqrt{2} /4$, leading to a naïve picture of four particles spanning the diagonal of the 
 lattice unit cell.  For $M=64$,  $a = 0.125$: There  is no evidence of lattice-induced structure (crosses) in the $45^o$ direction. However, a large peak (circle) at $k = 100.5$ implies structure in the horizontal (or vertical) direction with period $2\pi/k = 0.0625 = a/2$.

The take-away is that the order produced by the pins stems from the intricate details of packing of bidisperse particles among them.  There is every reason to assume $S(\vec{k})$ in lattice directions other than $0^o$ or $45^o$ will yield additional structural features, arising from the detailed way that particles pack among the pins.   For the two lattice directions studied above, local bond order and global spatial order don't transparently reinforce each other. For example,  no type of  $M=81$ bond pair shows a preference for $\theta = 0$,  while  Fig.  ~\ref{Sofk}a  shows its most dramatic peak in that lattice direction.

\subsection{Contact statistics}

\begin{figure}[h!]
	\centering
  \includegraphics[width=1.0\linewidth]{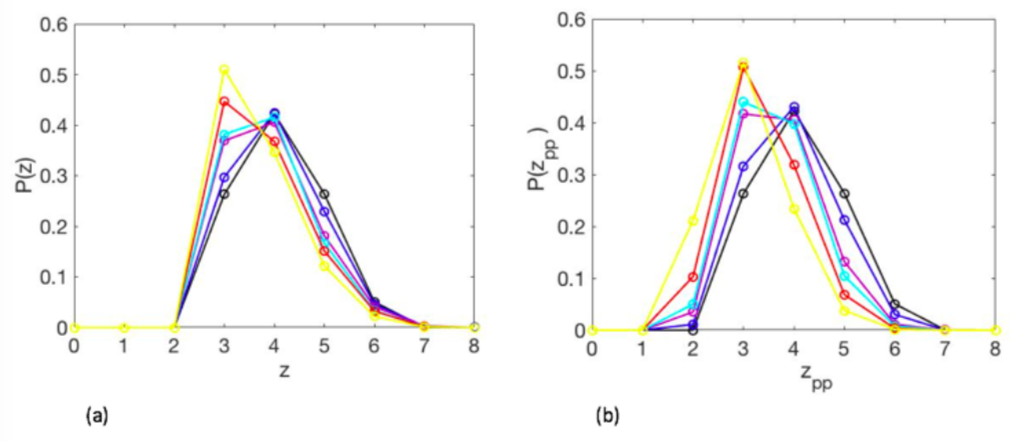}
  \caption{Probability that particle has $z$ contacts at $\phi = \phi_j(M)$  (see Table ~\ref{phijTable}).  Colors correspond to $M$ as in earlier figures.  a)  all contacts;  b) only particle-particle contacts.}
  \label{PZ}
\end{figure}

How do pins affect a particle's average contact number, $Z$?  In what follows,  the subscript ``pp"  denotes a contact between particles and ``pf" denotes contact between a particle and a fixed pin. It is obvious that $Z_{pp}$ is reduced by pins for a given value of $\phi$.  A particle stabilized by a pin might touch as few as two other particles and contribute to the rigid solid. It is not as obvious how $Z$ should vary with $M$. Fig. ~\ref{PZ} shows the probability $P$ for a particle to have $z$ contacts, evaluated at the configuration-averaged jamming threshold $\phi_j(M)$. Increasing $M$ at the jamming threshold will shift both distributions to the left, toward smaller number of contacts.   As Fig. ~\ref{Z} below shows, both $Z$ and $Z_{pp}$  {\em decrease} as pin density increases.  These results have technological consequences: A conductive jammed solid stabilized by pins may be expected to have a lower conductivity, higher individual bond strengths, lower yield strength, and different elastic moduli from its pin-free counterpart.  
\

The traditional Maxwell counting argument ~\cite{GoodrichPRL2012} asserts that frictionless, spherical particles require a minimum of $N_{B \ iso} = dN - qd + 1$ bonds. The second term in this definition arises from $d$ zero modes associated with global translations, while the third term ensures a positive bulk modulus.  Here,  $q=1$  without pins, but $q=0$ if even one pin is present, as our equilibration protocol breaks translational symmetry \cite{Graves2016}.  
Say that the total number of bonds is  $N_B = N_{pp} + N_{pf}$. The number of excess bonds between particles is found via a generalized isostaticity criterion ~\cite{Brito2013}:
\begin{equation}
N_{B \ excess} \equiv N_B - N_{B \  iso}  = N_{pp}+ N_{pf} - dN+ qd - 1 \ .
\label{Nexcess}
\end{equation}
Without pins,   $Z_{iso} = 2d - (2/N) = 2 N_{B \ iso}  $.  In the presence of pins it is no longer true that  one can write $Z_{iso}  \propto N_{B \ iso} $, as $N_{pp}$ bonds stabilize two particles, and $N_{pf}$ bonds stabilize only one.

Hyperstaticity, by which one means $N_{B} > N_{iso}$,  is expected in certain cases, such as frictional particles~\cite{Silbert2010}, or those with attractive interactions ~\cite{Lois2008, Koeze2020}. We see increasing hyperstaticity with $n_f$, reminiscent of previous work with frozen particles ~\cite{Brito2013} as well as in bidisperse mixtures in which the ratio of small to large radii is varied \cite{KoezeVagberg2016}. Unfortunately,  even at $n_f=0$, the number of excess bonds $N_{B excess,  \ 0}$  is greater than zero. This is a consequence of fixing $r_s/a$ for a set of initial conditions, thus averaging over configurations at various distances from their individual jamming points. (As one might hope, this fraction of excess bonds is independent of the system size, $N$.) Squares in Fig. ~\ref{Z} shows the difference between the number of excess bonds at finite $n_f$ and at $n_f=0$. 

Though $N_{B excess}$  increases with $n_f$,  the number of excess bonds {\em per pin} is found to decrease: 
\begin{equation}
N_{B excess}  - N_{B excess, \ 0} \sim n_f ^ \beta   ; \ \ \beta = 0.61 \pm 0.07
\label{NB}
\end{equation}
Eq. \ref{NB} shows an approximate square root dependence on pin number, which is reminiscent of a surface term.  Since both $Z$ and $Z_{pp}$ decrease with increasing pin density,  it is the number of bonds between particles and pins, $N_{pf}$, which increases in the viscinity of $\phi_j$.   The number of excess bonds reflect the interplay between the contact statistics (rising numbers of contacts between particles and pins) and the falling number of non-rattler particles at $\phi_j$.  

\begin{figure}[h!]
\centering
  \includegraphics[width=0.8\linewidth]{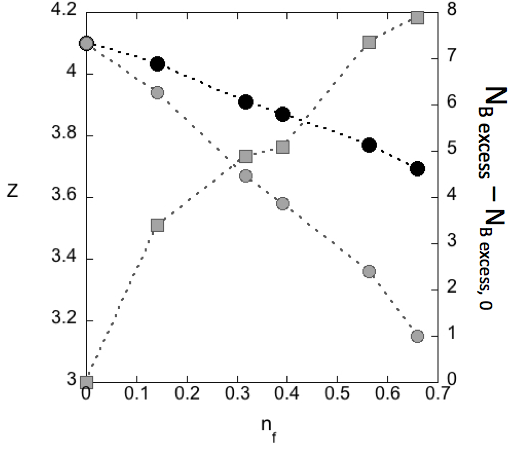}
  \caption{Typical number of contacts per particle.  Black circles:  All types of contacts, $z$,  grey circles: Particle-particle contacts, $z_{pp}$, grey squares:  
  $N_{B excess} - N_{B excess, \ 0}$ .  }
  \label{Z}
\end{figure}







\section{Conclusions}
 Introducing a square lattice of pin-like obstacles to systems of bidisperse particles provide a parsimonious route to jamming. The threshold $\phi_j$ decreases and fewer contacts are needed for stability as the density of pins increases. $\phi_j$ decreases linearly at low pin densities, and more steeply at higher densities.  This change in behavior occurs in a regime where the volume per particle is comparable to the volume per pin. There are additional, detailed changes in the structure of the jammed system as pin density increases.  The distribution of bond angles becomes increasingly anisotropic and we see a transition in a cubatic order parameter.  The bond angle distribution exhibits fourfold symmetry, consistent with the presence of pins in a square lattice, but with details that depend sensitively on the packing of bidisperse particles among pins of a given density. The presence of oscillations in the pair correlation function suggests long-range spatial ordering in the system. Peaks in the structure factor arise, locked to the spatial frequency of the pin lattice.   In general,  the axes along which there is long-range spatial order need not correspond to directions of preferred bond angles. This supports the notion that long-range, spatial ordering can be considered separately from local, bond ordering. Both are consequential when the pin separation is on the order of the particle size. 
\section*{Conflicts of interest}
There are no conflicts to declare.

\section*{Acknowledgements}
We thank Tristan Cates, Carl Goodrich, and M. Lisa Manning for invaluable technical contributions.   We  thank Cacey Bester, Peter Collings, Randall Kamien, Andrea Liu, Cynthia Olson Reichhardt, Daniel Sussman, Brian Utter, and Katharina Vollmayr-Lee for their comments and insights.  Acknowledgement is made to the donors of the American Chemical Society Petroleum Research Fund for partial support of this research and to the National Science Foundation: NSF Grant DMR-1905474. We are grateful to Swarthmore College's Provost, Division of Natural Sciences, and Individual Donors. A. L. Graves is grateful for a Michener Sabbatical Fellowship.  S. A. Ridout has been supported by PGS-D fellowship from NSERC and the Simons Foundation Cracking the Glass Problem Collaboration award No. 45494 to Andrea J. Liu.



\balance


\bibliography{rscv11} 
\bibliographystyle{rsc} 
\bibliographystyle{unsrt}

\end{document}